\documentclass[12pt]{iopart}
\usepackage{graphicx}
\def\be{\begin{equation}}
\def\ee{\end{equation}}
\def\bea{\begin{eqnarray}}
\def\eea{\end{eqnarray}}

\def\c{\cite}

\def\ov{\over}

\def\rs2r{{r_{s}\over 2r}}
\def\l2r2{{l^{2}\over r^{2}}}

\def\a2{{l^{2}\over a^{2}}}
\def\b2{{l^{2}\over b^{2}}}

\begin{document}

\jl{6}
\begin{center}
\title{Ponzano-Regge Model on Manifold with Torsion}

\author{T Vargas A.}
\address{Grupo de Astronom\'{\i}a-SPACE\\
Facultad de Ciencias F\'{\i}sicas\\
Universidad Nacional Mayor de San Marcos\\
Av. Universitaria s/n. Ciudad Universitaria, Lima 1.Lima-Peru}

\end{center}

\begin{abstract}

The connection between angular momentum in quantum mechanics and geometric objects is extended to manifold with torsion. First, we
notice the relation between the $6j$ symbol and Regge's discrete version of the action functional of Euclidean three dimensional
gravity with torsion, then consider the Ponzano and Regge asymptotic formula for the Wigner $6j$ symbol on this simplicial manifold
with torsion. In this approach, a three dimensional manifold $M$ is decomposed into a collection of tetrahedra, and it is assumed that
each tetrahedron is filled in with flat space and the torsion of $M$ is concentrated on the edges of the tetrahedron, the length of
the edge is chosen to be proportional to the length of the angular momentum vector in semiclassical limit. The Einstein-Hilbert action
is then a function of the angular momentum and the Burgers vector of dislocation, and it is given by summing the Regge action over all
tetrahedra in $M$. We also discuss the asymptotic approximation of the partition function and their relation to the Feynmann path
integral for simplicial manifold with torsion without cosmological constant.

\end{abstract}

%\pacs{04.20.-q; 04.50.+h; 04.60.Nc}

\maketitle

\section{Introduction}
The connection between angular momentum in quantum mechanics and geometric objects have been observed since long ago by Wigner~\cite{wi}.
In $1968$, Ponzano and Regge~\cite{pr} built a quantum gravity model in three dimensional Riemanian maniforld using the properties of some
invariants which can be obtained from the fundamental representation of the group $SU(2)$. Noting an interesting connection between the
$6j$-symbols for spin and the 3-dimensional Regge action for gravity~\cite{r}, they formulated the Feynman path integral for
three-dimensional simplicial quantum gravity in terms of a product of $6j$-symbols. Hasslacher and Perry~\cite{hp} not only elucidated
further details of this model and its relation to Penrose's spin netwoks~\cite{p}, but also discussed the possibility of spacetime foam
arising from this Ponzano and Regge model. At the early $90$'s Turaev and Viro~\cite{tv} proposed a generalization of this theory formulated
in terms of quantum $6j$-symbols and  showed that this generalization provided new 3-manifold invariants. Ooguri~\cite{o} demonstrated that
the Ponzano-Regge partition function is equivalent to Witten's 2+1 formulation of gravity~\cite{w} on closed orientable manifolds. Barrett
and Crane~\cite{bc} pointed out the relation between 6j-symbols and a discrete version of the Wheeler-de Witt equation, thereby giving
further insight into the relation of Ponzano-Regge theory to 3-dimensional gravity. Also in 2004, Freidel and Louapre~\cite{fl} considered
the Ponzano-Regge model in the context of spin foam quantization of three dimensional gravity coupled to quantum interacting spinning
particles.

On the other hand, there is an alternative approach to gravitation based on the Weitzenb\"ock geometry~\c{we}, a manifold with torsion
without curvature, which is so called teleparallel gravity. In this theory, gravitation is attributed to torsion~\cite{xs}, which plays
the role of a force~\cite{pe}, and the curvature tensor vanishes identically. As is well known, at least in the absence of spinor fields,
teleparallel gravity is equivalent to general relativity. In this paper, relying upon this equivalence, first we will obtain the relation
between the $6j$ symbol and Regge's discrete version of the action functional of Euclidean three dimensional gravity with torsion without
cosmological constant, then considere the Ponzano and Regge asymptotic formula for the Wigner $6j$ symbol on this simplicial manifold
with torsion. In this approach, a three dimensional Weitzenb\"ock manifold M is decomposed into a collection of tetrahedra, and it is
assumed that each tetrahedron is filled in with flat space and the torsion of M is concentrated on the edges of the tetrahedron, the length
of the edge is chosen to be proportional to the length of the angular momentum vector in semiclassical limit. The Einstein-Hilbert action is
then a function of the angular momentum and the Burgers vector of dislocation, and it is given by summing the Regge action over all
tetrahedra in $M$.

It is interesting to note that there is a relation between an {\it angle} $\varepsilon_i$, which is formed by the outward normals of
two faces sharing the $i-$th edge and the {\it distance} $b_i$ through which a vector is {\it translated} from its original position
when parallel transported around a small loop of area $\Sigma^{\ast}_i$. The partition function is invariant under a refinement of a
simplicial decomposition, more specifically, it is invariant the $1-4$ and $2-3$ Alexander moves and its asymptotic approximation can
be interpreted as the path integral formulation of three dimensional simplicial Euclidean manifold with torsion, provided the number of
edges and vertices in the simplical manifold becomes very large and in the sum over edges the large values dominate.

It is noteworthy to remark that, in a vacuum three-dimensional simplicial manifold with torsion, the torsion tensor is localized in
one-dimensional dislocation line, called hinges. Torsion can then be detected by measuring the dislocation in relation to the initial
position of a vector, as a result of the parallel transport along a small loop encircling the dislocation line (hinge), where torsion is
concentrated. When torsion is present, it is detected a dislocation parallel to this hinge, and this dislocation is measured by the Burgers
vector $b_d$. The resulting simplicial action is considered as the teleparallel equivalent of the simplicial Einstein's action of general
relativity.  We will proceed according to the following scheme. In section~2, we review the main features of teleparallel gravity. In section~3, we obtain
the simplicial torsion and the discrete action. In section~4, we considere the Ponzano-Regge asymptotic formula for the Wigner $6j$ symbol
on this simplicial manifold with torsion. Discussions and conclusions are presented in section~5.

\section{Teleparallel Equivalent of General Relativity}

It is well known that curvature, according to general relativity, is used to geometrize the gravitational interaction. On the other hand,
teleparallelism attributes gravitation to torsion, but in this case torsion accounts for gravitation not by geometrizing the interaction,
but by acting as a force~\c{pe}. This means that in the teleparallel equivalent of general relativity, instead  of geodesics, there are
force equations quite analogous to the Lorentz force equation of electrodynamics.

A nontrivial tetrad field induces on spacetime a teleparallel structure which is directly related to the presence of the gravitational
field. In this case, tensor and local tangent indices\footnote{The greek alphabet ($\mu$, $\nu$, $\rho$,~$\cdots=1,2,3$) will be used to
denote tensor indices and the latin alphabet ($a$, $b$, $c$, ~$\cdots=1,2,3$) to denote local tangent space indices, whose metric tensor
is chosen to be $\eta_{ab} = \mbox{diag} (+1, +1, +1)$. Furthermore, we will use units in which $c =1$.} can be changed into each other
with the use of a tetrad field $h^{a}{}_{\mu}$. A nontrivial triad field can be used to define the linear Weitzenb\"ock connection
\be
\Gamma^{\sigma}{}_{\mu \nu} = h_a{}^\sigma \partial_\nu h^a{}_\mu,
\label{car}
\ee
a connection presenting torsion, but no curvature. It parallel transports the tetrad itself:
\be
{\nabla}_\nu \; h^{a}{}_{\mu}
\equiv \partial_\nu h^{a}{}_{\mu} - \Gamma^{\rho}{}_{\mu \nu}
\, h^{a}{}_{\rho} = 0.
\label{weitz}
\ee
The Weitzenb\"ock connection satisfies the relation
\be
{\Gamma}^{\sigma}{}_{\mu \nu} = {\stackrel{\circ}{\Gamma}}{}^{\sigma}{}_{\mu
\nu} + {K}^{\sigma}{}_{\mu \nu},
\label{rel}
\ee
where
\be
{\stackrel{\circ}{\Gamma}}{}^{\sigma}{}_{\mu \nu} = \frac{1}{2}
g^{\sigma \rho} \left[ \partial_{\mu} g_{\rho \nu} + \partial_{\nu}
g_{\rho \mu} - \partial_{\rho} g_{\mu \nu} \right]
\label{lci}
\ee
is the Levi--Civita connection of the metric
\be
g_{\mu \nu} = \eta_{a b} \; h^a{}_\mu \; h^b{}_\nu,
\label{gmn}
\ee
and
\be {K}^{\sigma}{}_{\mu \nu} = \frac{1}{2}
\left[T_{\mu}{}^{\sigma}{}_{\nu} +T_{\nu}{}^{\sigma}{}_{\mu}-T^{\sigma}{}_{\mu \nu}\right]
\label{conto}
\ee
is the contorsion tensor, with
\be
T^\sigma{}_{\mu \nu} =
\Gamma^{\sigma}{}_{\nu \mu} - \Gamma^ {\sigma}{}_{\mu \nu} \;  \label{tor}
\ee
the torsion of the Weitzenb\"ock connection.

Since here we are interested in Euclidean three dimensional teleparallel gravity, we define the action on the manifold with
torsion as
\be
S=\int d^{3}x \,L_{T} \, .
\label{ta}
\ee
Here $L_{T}$ is the teleparallel gravitational lagrangian given by
\be
L_{T} = \frac{h}{16 \pi G}\left[{1\ov 4} T^{\rho}{}_{\mu \nu}T_{\rho}{}^{\mu \nu} +
{1\ov 2} T^{\rho}{}_{\mu \nu}T^{\nu \mu}{}_{\rho} - T_{\rho \mu}{}^{\rho}T^{\nu \mu}{}_{\nu}
\right],
\label{la}
\ee
with $h = \det(h^{a}{}_{\mu})$. These are all necessary information we need about the teleparallel gravity, since only the discrete
version of the action (\ref{ta}) is required in both asymptotic approximations of $6j$ symbol and the partition function.

\section{Discrete torsion and simplicial action}

In order to gain further insight into the simplicial manifold with torsion we begin by reviewing the Regge calculus on which the derivation
of simplicial teleparallel action is based~\cite{pv}. Similarly to the Regge construction of the simplicial manifold of general relativity,
we assume that the usual continuous spacetime manifold with torsion can be viewed as the limit of a suitable sequence of discrete lattices
composed of an increasing number of smaller an smaller simplices. In general, the Weitzenb\"ock manifold, which is the stage set of
teleparallel gravity, is approximated by a $D$-dimensional polyhedra $M^D$. In this approach, the interior of each simplex is assumed to be
flat, and this flat $D$-simplices are joined together at the $D$-hedral faces of their boundaries. The torsion turns out to be localized in
the $D-2$-dimensional dislocation simplices (hinges) of the lattice, and the link lengths $l$  between any pair of vertices serve as
independent variables.

To begin with, let us take a bundle of parallel dislocations (hinges) in $M^3$. We make the assumption that the torsion induced by the
dislocations is small, so that we may regard $M^3$ as approximately euclidian. Let ${\bf U}$ be a unity vector parallel to the dislocations.
We test for the presence of torsion by carrying a vector ${\bf A}$ around a small loop of area vector ${\bf S}= S{\bf n}$, with $S$ the
area and ${\bf n}$ a unity vector normal to the surface. At the end of the test, if torsion is nonvanishing, ${\bf A}$ is found to have
translated from the original position, along ${\bf U}$, by the length ${\bf B} = N {\bf b}$, where $N$ is the number of dislocations
entangled by the loop, and ${\bf b}$ is the Burgers vector, which is a vector that gives both the length and direction of the closure
failure for every dislocation. In $M^4$, the flux of dislocation lines through the loop of area $S^{\alpha \beta}$ is
\[
\Phi = \rho \left({\bf U}{\bf S} \right) = {1\over2} \, \rho_{\alpha \beta} \, S^{\alpha \beta},
\]
where $\rho$ is the density of dislocation passing through the loop, and $\rho_{\alpha \beta} =
\rho \, U_{\alpha \beta}$, with $U^{\alpha \beta}$ a unity antisymmetric tensor satisfying
$U_{\alpha \beta}U^{\alpha \beta} = 2$. This means that we  can endow the polyhedra more densely with hinges in a region of high torsion
than in region of low torsion. The closure failure is then found to be
\be
B_{\mu} = {1\over2} \, \rho_{\alpha \beta} \, S^{\alpha \beta}\,b_{\mu}.
\ee
However, we know from differential geometry that, in the presence of torsion,
infinitesimal parallelograms in spacetime do not close, the closure failure being equal to
\be
B_{\mu}=T_{\mu \nu \sigma} \, {S}^{\nu \sigma}.
\label{bts}
\ee
By comparing the last two equations we see that
\be
T_{\mu \alpha \beta} = \frac{1}{2} \, \rho_{\alpha \beta} \,b_{\mu} \equiv
\frac{1}{2} \, \rho \, U_{\alpha \beta} \,b_{\mu}.
\label{tordef}
\ee

On the other hand, it has already been shown~\c{t} that the torsion singularity takes the form of a conical  singularity. Consequently, the
dislocation from the original position that occurs when a vector is parallel transported around a small loop encircling a given bone is
independent of the area of the loop. Furthermore, the dislocation has the next main characteristics: The length of the dislocation line in
three-dimensions, and the area of the triangle in four-dimensions. Therefore, there is a natural unique volume associated with each
dislocation. To define this volume, there is a well-known procedure in which a  dual lattice is constructed for any given
lattice~\c{ch,mi}. This involves constructing polyhedral cells around each vertex, known in the literature as Voronoi polygon, in such a
way that the polygon around each particular vertex contains all points which are nearer to that vertex than to any other vertex. The
boundary of the Voronoi polygon is always perpendicular to the edges emanating from the vertex, and each corner of the Voronoi polygon
lies at the circuncentre of any of the simplices of the Delaunay geometry, which shares the dislocation (bone) (see Fig.~1).

%%%%%%%%%%%%%%%%%%%
\begin{figure}
\begin{center}
\scalebox{0.35}[0.35]{\includegraphics{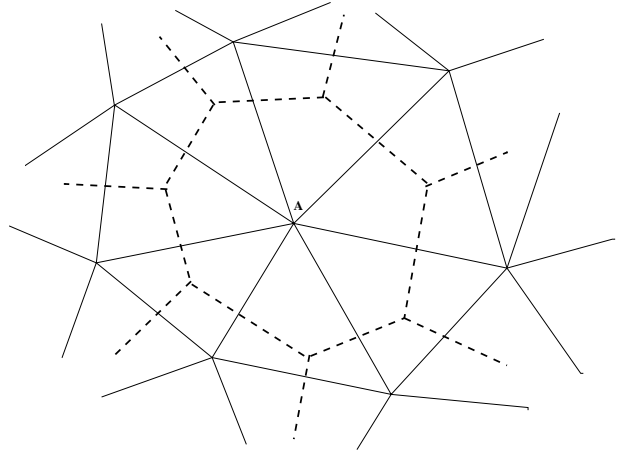}}
\end{center}
\caption{The two-dimensional Voronoi polygon (dashed line) for a particular vertex $A$. Each
corner of the Voronoi polygon lies at the circuncentre of any of the triangles of the Delaunay
geometry (solid line).}
\label{1}
\end{figure}
%%%%%%%%%%%%%%%%%%%

By construction, the Voronoi polygon is orthogonal to the bone. If we parallel transport a vector around the perimeter of a Voronoi polygon
of area ${\Sigma^{\ast}_{d}}$, it will traverse the flat geometry of the interior of each one of the simplices sharing the bone, and will
return dislocated from its original position in a plane parallel to the bone by a length $b_{\mu}$. According to this construction, and
relying on the definition (\ref{tordef}), the torsion due to each dislocation can be expressed by (see Fig.~2):
\be
\mbox{(Torsion)} = \frac{\mbox{(Distance the vector is translated)}}{\mbox{(Area
circumnavigated)}}.
\ee
On the other hand, it is well known that the Riemann scalar is proportional to the Gauss curvature, and proportionality constant depends
on the dimension $D$ of the lattice geometry~\c{mi}. In similar way, we define the simplicial torsion due to each dislocation by
\be
T_{(d)\mu \nu \rho} = \sqrt{D(D-1)} \ \; \frac{b_{(d)\mu}U_{(d)\nu \rho}}{{\Sigma^{\ast}_{d}}}
\equiv \sqrt{6} \ \; \frac{b_{(d)\mu}U_{(d)\nu \rho}}{{\Sigma^{\ast}_{d}}}.
\label{T}
\ee
The reason for the square root is that, as is well known from teleparallel gravity, the Riemann curvature tensor is proportional to a
combination of squared torsion tensors.

As we have said, the vector returns translated from its original position in a plane parallel to the hinge by a length $b^{\mu}$. Let us
then analyze the translational group acting on it. As we know, the interior of each block is flat, so the infinitesimal translation in
these blocks is given by
\be
T(\delta b) = I - i \; \delta b^{a} \, P_{a},
\ee
where $I$ is the unity matrix, $\delta b^{a}$ are the components of an arbitrarily small three-dimensional Burgers (displacement) vector,
and $P_{a} = i \partial_{a}$ are the translation generators. In the presence of dislocations, and using the tetrad $h^{a}{}_{\mu}$, this
infinitesimal translation parallel to the hinge becomes
\be
T(\delta b) = I - i \; \delta b^{\mu} \, h^{a}{}_{\mu} \, P_{a},
\ee
so that a finite translation will be represented by the group element
\be
T(b) = \exp \left[ -i \, b^{\mu} \, h^{a}{}_{\mu} \, P_{a} \right].
\ee
On the other hand, the contour integral of the Burgers vector --- which measures how much the infinitesimal closed contour $\Gamma$ spanning a surface element
$d \Sigma^{\ast \alpha \beta}$ fails to close in the presence of hinge --- by using
Eq.~(\ref{bts}), is seen to be~\c{kle}
\be
b^{\mu}= \oint\limits_{\Gamma} T^{\mu}{}_{\alpha \beta} \, d \Sigma^{\ast \alpha \beta}.
\ee
Therefore, the group element of translations due to torsion turns out to be
\be
T(b) = \exp \left[-i\oint\limits_{\Gamma} P_{a} \, T^{a}{}_{\alpha \beta} \,
d \Sigma^{\ast \alpha \beta} \right].
\ee
%%%%%%%%%%%%%%%%%%%
\begin{figure}
\begin{center}
\scalebox{0.35}[0.35]{\includegraphics{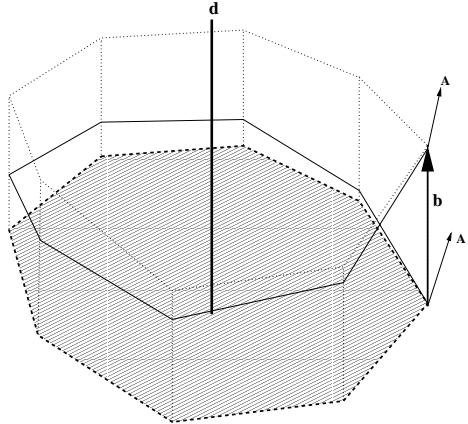}}
\end{center}
\caption{The parallel transport of a vector $A$ along the perimeter of a Voronoi polygon (dashed line) around the dislocation line in
three-dimension. The vector returns translated in a plane parallel to the dislocation $d$ by the length $b$.}
\label{2}
\end{figure}
%%%%%%%%%%%%%%%%%%%

The $D$-volume $\Omega_d$ associated with each dislocation, as described above, is proportional to the product of $\Sigma_d$, the
$D-2$-dimensional volume of the dislocation, and $\Sigma^{\ast}_d$, the area of the Voronoi polygon~\c{mi}:
\be
\Omega_d \equiv \frac{2}{D(D-1)} \; \Sigma_d \, {\Sigma^{\ast}_d} =
\frac{1}{3} \; \Sigma_d \, {\Sigma^{\ast}_d}.
\ee
The invariant volume element $h \, d^{3}x$, therefore, is represented by $\Omega_d$, and we have the following relation,
\be
\int h \, d^{3}x \Longrightarrow \sum_{\rm dis}\Omega_d = \frac{1}{3} \;
\sum_{\rm dis} \Sigma_d \, {\Sigma^{\ast}_d},
\ee
where the sum is made over all dislocations. We are ready then for constructing the simplicial action. Let us take the
lagrangian (\ref{la}) of teleparallel gravity, whose terms are proportional to the square of the torsion tensor, and substitute torsion as
given by Eq.~(\ref{T}). For the first term, we obtain
\be
T_{(d)}^{\mu \nu \rho} \, T_{(d)\mu \nu \rho} =
6 \left(\frac{1}{{\Sigma^{\ast}_d}} \right)^{2}b^{\mu}_{(d)} \, b_{(d) \mu}.
\ee
Writing the other two terms in a similar way, the simplicial teleparallel action will be
\be
S_{T} = \frac{1}{16 \pi G}\sum_{\rm dis} \left(\frac{b^2_d}{{\Sigma^{\ast}_d}}\right)l_d,
\label{tpa}
\ee
where $b^{2}_{d}$ denotes the projected Burgers vector parallel to the hinge,
and $l_d$ is the dislocation line, or hinge.

\section{Ponzano-Regge model}

Ponzano and Regge~\cite{pr} studied a model in which the simplicial bloks of three-dimensional Riemanian manifold are $3$-dimensional
tetrahedra. Each edge of a tetrahedron is labelled by a half integer $j$, corresponding to the $(2j+1)$-dimensional fundamental
representation of the group $SU(2)$, such that $\sqrt{j(j+1)}\hbar \approx (j+\frac{1}{2})\hbar$ for large $j$ is the descritized length
of that edge. A tetrahedron with four vertices, six edges and four triangular faces is then the natural geometric representation of the
recoupling coefficients between four angular momenta (see Fig. 3).
%%%%%%%%%%%%%%%%%%%
\begin{figure}
\begin{center}
\scalebox{0.35}[0.35]{\includegraphics{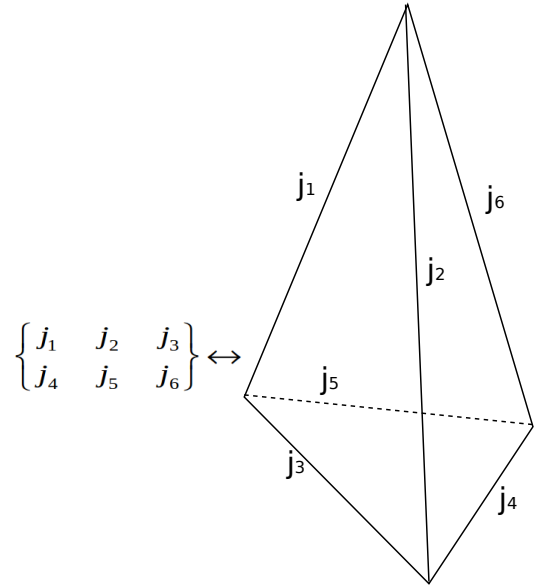}}
\end{center}
\caption{Geometric representation of $6j$-coefficients.}
\label{3}
\end{figure}
%%%%%%%%%%%%%%%%%%%
There is thus a one-to-one correspondence betwen the number of edges of tetrahedron and the number of arguments of the $6j$-symbol, namely
\be
l_i =(j_i +\frac{1}{2})\hbar\, , \,i=1,2,...,6.
\label{lj}
\ee
These lengths must satisfy the triangle inequalities corresponding to the triangular faces of the tetrahedron, then $6j$-symbol is
only defined when the values of the angular momenta triades which correspond to the edge lengths around a face astisfy the triangle
inequalities,$\mid{j_1-j_2}|\leq j_3\leq j_1+j_2$. This implies that the edge lenghts around a face satisfy the triangular inequalities,
up to additional terms of $\pm \frac{1}{2}$: though four momenta triads satisfy triangle inequalities, the same triades shifted by
$1/2$ need not. In that case the $6j$ symbol is said to be classically forbidden, and it is exponentially suppressed at large $j_i$. Also,
$j_1$, $j_2$, $j_3$ are required to satisfy $j_1+j_2+j_3=$integer, for each face. These inequalities for the angular momentum guarantees
that the edges $l_1$, $l_2$, $l_3$ of tetrahedron form a closed triangle af non-zero surface area: $\mid{l_1-l_2}| < l_3 < l_1+l_2$. If
these inequalities are not satisfied, the value of the $6j$-symbol is defined to be zero.

Ponzano and Regge main porpuse was to evaluate the asymptotic limit of $6j$-symbol when all six arguments $j_1$, $j_2,......$, $j_6$
are large or when $j_i$ lie in the classically allowed region, and the asymptotic formula obtained was
\be
\left\{\eqalign{j_1 \,\,\, j_2 \,\, j_3 \cr j_4 \,\,\, j_5 \,\,\, j_6 \cr}\right\}\simeq
\sqrt{\frac{\hbar^{3}}{12\pi V(j)}}\cos\left[\frac{1}{\hbar}\sum_{i=1}^{6}
(j_i +\frac{1}{2})\varepsilon_i +\frac{\pi}{4}\right]\, ,
\label{fase}
\ee
where $\varepsilon_i$ is the interior angle between the outward normals of the tetrahedral faces sharing the $l_{i}^{th}$ edge (see Fig. 4).
%%%%%%%%%%%%%%%%%%%
\begin{figure}
\begin{center}
\scalebox{0.35}[0.35]{\includegraphics{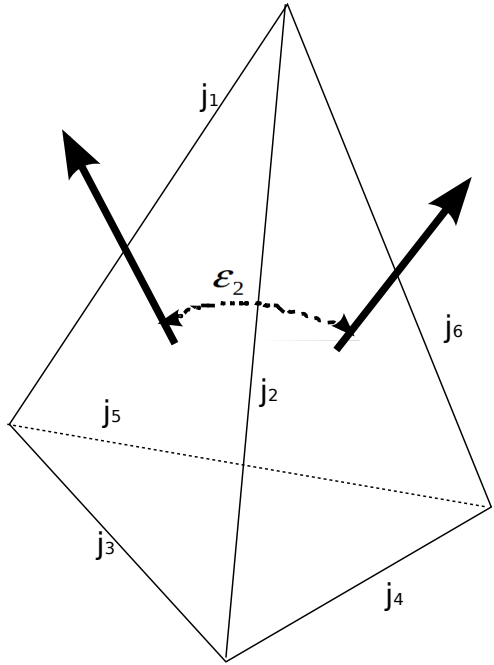}}
\end{center}
\caption{$\varepsilon_i$ is the angle between the outward normals of the tetrahedral faces which have
the edge $j_{i}^{th}$ in common.}
\label{4}
\end{figure}
%%%%%%%%%%%%%%%%%%%

Because we are interested in the large $j_i$ limit, the shifts by $1/2$ may be neglected and the square of the tetrahedral volume $V(j)$ can
be found from the Cayley formula:
\be
V(j)^{2}=\frac{1}{288}\left|\,\matrix{
0&j_{4}^2\hfill&j_{5}^2\hfill &j_{6}^2&\hfill 1\cr
j_{4}^2 &0\hfill&j_{3}^2\hfill &j_{2}^2&\hfill 1 \cr
j_{5}^2&j_{3}^2 &\hfill 0&j_{1}^2 &\hfill 1 \cr
j_{6}^2&j_{2}^2\hfill&j_{1}^2&\hfill 0&\hfill 1 \cr
1&1\hfill&1\hfill &1\hfill &0 \hfill \cr }\right| \, .
\ee
In order to obtain a meaninful non-imaginary result from equation (\ref{fase}) the tetrahedral volume must be real. It means that, tetrahedra
with non-negative $V^2$ can be embedded in a three-dimensional Euclidean space, while the same can not be done with tetrahedra with
negative $V^2$. However, the $V^2 < 0$ tetrahedra can be embedded in a three-dimensional Lorentzian space~\cite{bf}. Let us remark that
the Ponzano and Regge asymptotic formula (\ref{fase}) were later elaborated notably by Schulten and Gordon~\cite{sg} and the first rigorous
proof of it has been obtained by Roberts~\cite{ro} using geometric quantization.

On the other hand, the simplicial three-dimensional Einstein-Hilbert action or Regge action of general relativity is~\cite{r}
\be
S_{R} = \frac{1}{8 \pi G} \sum_{\rm hinge} \; \varepsilon_h \; l_h,
\label{rga}
\ee
where $\varepsilon_h$ is the deficit angle associated to each hinge, which is directly related to the curvature of spacetime, and for the
tetrahedron using (\ref{lj}) it is given by
\be
S_{R}=\frac{1}{8 \pi G}\sum_{i=1}^{6}l_i\varepsilon_i=
\frac{1}{8 \pi G}\sum_{i=1}^{6}(j_i +\frac{1}{2})\varepsilon_i \, ,
\label{ram}
\ee
which means that, the gravitational contribution at each edge of the tetrahedron is $l_i\varepsilon_i=(j_i +\frac{1}{2})\varepsilon_i$.
Therefore, the semiclasssical approximation (\ref{fase}) is equal to the cosine of the Regge action for a single tetrahedron up to a
constant factor and a phase shift: $\cos\left(\frac{S_{R}}{\hbar} +\frac{\pi}{4}\right)$.

We now turn to the Ponzano-Regge model on manifold with torsion. In a previous section we have seen that the usual continuous spacetime
manifold with torsion can be viewed as the limit of a suitable sequence of discrete lattices composed of an increasing number of smaller
an smaller simplices, and in a vacuum three-dimensional case, the torsion tensor is localized in one-dimensional dislocation line $l_i$,
called hinges. Also we have demostrated that when torsion is present, it is detected a dislocation parallel to this hinge, and this
dislocation is measured by the Burgers vector $b_d$. From these set of $l_i$, let us choose six of them in such a way that they must
satisfy triangle inequalities. More precisely, let $l_1$, $l_2$, $l_3$, $l_4$, $l_5$, $l_6$ be non-negative integers and an unordered
triades of this family of dislocation lines written as $(l_i, l_j, l_k)$ with $i,j,k=1,2,...,6$ and $i\neq j \neq k$, is said to be
admissible if the triangular inequalities $\mid{l_j-l_k}| < l_i < l_j+l_k$ are met. These admissible $l_i$ are the edge lengths of the
tetrahedron and also they completely characterizes it in Euclidean $3$-space: the triads
$(l_1,l_2,l_3)$, $(l_3,l_4,l_5)$, $(l_1,l_5,l_6)$ and $(l_2,l_6,l_4)$ form the four faces of the tetrahedron and the following determinant
\be
V(l)^{2}=\frac{1}{288}\left|\,\matrix{
0&l_{4}^2\hfill&l_{5}^2\hfill &l_{6}^2&\hfill 1\cr
l_{4}^2 &0\hfill&l_{3}^2\hfill &l_{2}^2&\hfill 1 \cr
l_{5}^2&l_{3}^2 &\hfill 0&l_{1}^2 &\hfill 1 \cr
l_{6}^2&l_{2}^2\hfill&l_{1}^2&\hfill 0&\hfill 1 \cr
1&1\hfill&1\hfill &1\hfill &0 \hfill \cr }\right|
\ee
must be positive.

Then, the simplicial teleparallel action (\ref{tpa}) for the tetrahedron reduces to
\be
S_{T} = \frac{1}{16\pi G}\sum_{i=1}^{6}l_i\left(\frac{b^2_i}{{\Sigma^{\ast}_i}}\right) \, ,
\ee
where $l_{i}$, $b_i$ are the edge length and the closure failure or gap at the edge correspondingly. ${\Sigma^{\ast}_{i}}$ is the area of
a Voronoi polygon which is orthogonal to the edge. As mentioned in the introduction of this section, the tetrahedron is the natural
geometric representation of the $6j$-symbol, and there is one-to-one correspondence between the number of tetrahedral edges and the number
of arguments of the $6j$-symbol given by (\ref{lj}). Then, the Regge action may be re-expressed as
the sum of the gravitational contribution from each edge of the tetrahedron:
\be
S_{T} = \frac{1}{16\pi G}\sum_{i=1}^{6}(j_i +\frac{1}{2})\left(\frac{b^2_i}{{\Sigma^{\ast}_i}}\right) \, .
\label{prt}
\ee
If we have a complex of tetrahedra with $N$ internal edges, then the action would be written as
\be
S_{T} =\frac{1}{16\pi G}\sum_{i=1}^{N}(j_i +\frac{1}{2})\left(\frac{b^2_i}{{\Sigma^{\ast}_i}}\right) \, .
\label{tam}
\ee
The discrete action is then a function of the angular momentum, the Burgers vector of dislocation and the area of Voronoi polygon.

Consequently, in this approach we can approximate a three dimensional manifold with torsion $M$ by a large collection of tetrahedra glued
together. If we assume that each tetrahedron is filled in with flat space and the torsion of $M$ is concentrated on the edges of the
tetrahedron, and since the length of the edge is chosen to be proportional to the length of the angular momentum vector in semi classical
limit, a metric $g_{\mu \nu}$ which is related to triad on $M$ by (\ref{gmn}) is specified once the length $(j_i +\frac{1}{2})$ of each
edge is fixed. The Euclidean Einstein-Hilbert action given by (\ref{ta}) is then a function of the angular momentum on the edges and is
given by summing the simplicial action (\ref{prt}) over all the tetrahedra in $M$.

Having identified the edges $l_{i}$ of the tetrahedron where the torsion is concentrated with the angular momenta, the asymptotic form
of $6j$ symbol for large values of $j_i$ is given by
\be
\left\{\eqalign{j_1 \,\,\, j_2 \,\, j_3 \cr j_4 \,\,\, j_5 \,\,\, j_6 \cr}\right\}\simeq
\sqrt{\frac{\hbar^{3}}{12\pi  V(j)}}\cos\left[\frac{1}{8\pi G\hbar}\sum_{i=1}^{6}(j_i +
\frac{1}{2})\left(\frac{b^2_i}{{\Sigma^{\ast}_i}}\right) + \frac{\pi}{4}\right]\,.
\label{prasymp}
\ee
This is the Ponzano and Regge asymptotic formula for the Wigner $6j$ symbol on simplicial manifold with torsion. Here $V(j)$ is the three
dimensional volume of the tetrahedron and $b_i$ is the Burgers vector which gives both the length and
direction of the closure failure or gap for every dislocation in the tetrahedron corresponding to the edge $j_{i}+\frac{1}{2}$.
${\Sigma^{\ast}_{i}}$ is the area of a Voronoi polygon which is located in the plane perpendicular to a edge of tetrahedron.

Comparing right hand side of the asymptotic equations (\ref{fase}) and (\ref{prasymp}), we obtain a relation between the
{\it angle} $\varepsilon_i$, which is formed by the outward normals of two faces sharing the $i-$th edge and the {\it distance} $b_i$
through which a vector is {\it translated} from its original position when parallel transported around a small loop of
area $\Sigma^{\ast}_i$ perpendicular to $l_{i}$.

In this way, in closed three dimensional Euclidean manifold with torsion $M$ we have introduced a tetrahedral triangulation, and
have associated $6j$ symbol to each tetrahedron. Then, the set of vertices, edges, faces and tetrahedra with various $6j$ symbols
associated to the tetrahedra defines the simplicial decomposition of $M$. Following Ponzano and Regge, we define a partition function
by summing over all possible edge lengths simillar to Regge calculus and by taking the product of the $6j$ symbols over all fixed
number of tetrahedra and connectivity of the simplicial manifold:
\be
Z_{PR}(M)= \lim_{L\to\infty}\sum_{j\leq L}\prod_{vertices}\Lambda(L)^{-1}\prod_{edges}(2j+1)
\prod_{tetrahedra}(-1)^{\sum_i j_i}\left\{\eqalign{j_1 \,\,\, j_2 \,\, j_3 \cr j_4 \,\,\, j_5 \,\,\, j_6 \cr}\right\} \,
\label{parfunc}
\ee
where $L$ is a non-negative integer or half-integer cut off and the factor $\Lambda(L)^{-1}$ per each vertex was introduced to regulate divergences and is defined as
\be
\Lambda(L) = \sum_{p=0,\frac{1}{2},1,...,L}\left(2p+1\right)^2 \,
\ee
which behaves as $\Lambda(L)\sim \frac{4L^3}{3}$ in the limit $L\longrightarrow \infty$.

An important characteristic of this partition function is that it is independent of the way the interior of the manifold is
triangulated and it is invariant under a refinement of a simplicial decomposition, more specifically, it is invariant under the
following $1-4$ and $2-3$ Alexander moves: One can decompose a tetrahedron into four tetrahedra by adding a new site in the middle
of the original tetrahedron and by inserting an edge into two tetrahedra sharing the trinagle. Repeated applications of these moves
yield a finer and finer mesh without altering the value of $Z_{PR}(M)$. Thus the partition function defines a topological invariant
of the manifold $M$ and it is expected to be a topologiacl field theory~\cite{ka}.

Let us remark that the sum of contributions to $S_{T}$ in (\ref{tpa}) from all tetrahedra in a tesselation approaches a value
proportional to the action of teleparallel gravity, S in (\ref{ta}), provided the number of edges and vertices in the simplical
manifold becomes very large:
\be
\lim_{N\to\infty}\sum_{j=i}^{N}(j_i +\frac{1}{2})\left(\frac{b^2_i}{{\Sigma^{\ast}_i}}\right) \simeq 16\pi G S =\int d^{3}x \,L_{T} \, .
\label{simcont}
\ee
Comparing (\ref{parfunc}) and (\ref{simcont}) we see that the partition function $Z_{PR}(M)$ can be interpreted as the path integral
formulation of three dimensional Euclidean teleparalled gravity on a lattice. As it was remarked in~\cite{will}, a given $6j$ symbol
is proportional to the path integral amplitude for the associated tetrahedron, so the product of all $6j$ symbols is equivalent to
the path integral amplitude for a given simplicial geometry: If in the sum over edges the large values dominate, then the sum over
$j_{i}$ is replaced by an integral and the asymptotic value for $6j$ symbols is used, $Z_{PR}(M)$ contains a term proportional to
\be
\int\prod_{i} dj_{i}(2j_{i}+1)\prod_{tetrahedra \,\, p}\sqrt{\frac{\hbar^{3}}{12\pi  V_{p}}}
( e^{iS_{T}}+e^{-iS_{T}})=\int\prod_{i} d\mu(j_{i})e^{iS_{T}} \,
\label{pi}
\ee
which looks like a Feynmann path integral with the Regge action in three dimension, and with other terms
contributing to the measure of the integral. However this interpretation suffer from two problems: The integrals is over the form
$e^{iS_{T}}$ rather than $e^{-S_{T}}$ although we are considering Euclidean three dimensional spacetime, and the justification of
associating various interference terms to measure by hand. Despite these difficulties, it seems natural to expect that this asymptotic
approximation of the partition function leads to Feynmann path integral for simplicail manifold with torsion and the $6j$ symbol
appears to be related to semiclassical quantum gravity.

\section{Final Remarks}

In this paper we have considered the connection between angular momentum in quantum mechanics and geometric objects, namely the relation
between angular momentum and tetrahedra on manifold with torsion without the cosmological term. First, we noticed the relation between
the $6j$ symbol and Regge's discrete version of the action functional of Euclidean three dimensional gravity with torsion given by
(\ref{prt}). Then we considered the Ponzano and Regge asymptotic formula for the Wigner $6j$ symbol on this simplicial manifold with
torsion (\ref{prasymp}). Let us remark that in this approach, a three dimensional manifold $M$ is decomposed into a collection of
tetrahedra, and it is assumed that each tetrahedron is filled in with flat space and the torsion of $M$ is concentrated on the edges of
the tetrahedron, the length of the edge is chosen to be proportional to the length of the angular momentum vector in semiclassical limit.
The Einstein-Hilbert action on this manifold is then a function of the angular momentum and the Burgers vector of dislocation, and it is
given by summing the Regge action over all tetrahedra in $M$ (\ref{tam}).

Following Ponzano and Regge~\cite{pr}, we defined a partition function (\ref{parfunc}) by summing over all possible edge lengths simillar
to Regge calculus and by taking the product of the $6j$ symbols over all fixed number of tetrahedra and connectivity of the simplicial
manifold, and in order to regulate the divergences by a cut off we have introduced a non-negative integer or half-integer parameter $L$.
Although the partition function is finite in simple examples, it diverges in general cases since the set of irreducible representations
of $SU(2)$ is infinite and the partition function is often a sum of an infinite number of terms, as explained in~\cite{bn}. A regularization
of this infinite sum and its relation to the cosmological constant was provided by the Turaev$-$Viro model~\cite{tv} and~\cite{mt} by replacing
the Lie group $SU(2)$ by its quantum deformation $U_{q}(sl_{2}(C))$, which has only a finite number of representation. This q-deformed
Ponzano Regge model on manifold with torsion will be given elsewhere.

The asymptotic approximation of the partition function can be interpreted as the path integral formulation of three dimensional Euclidean
teleparalled gravity on a lattice, since a given $6j$ symbol is proportional to the path integral amplitude for the associated tetrahedron,
so the product of all $6j$ symbols is equivalent to the path integral amplitude for a given simplicial geometry. This interpretation is
possible because in the sum over edges only the large values dominate and the number of edges and vertices in the simplical manifold
is very large. Consiquently, equation (\ref{pi}) looks like a Feynmann path integral for simplicail manifold with torsion with the
Regge action in three dimension and the $6j$ symbol appears to be related to semiclassical quantum gravity.

%\ack
%The authors would like to thank Yu. N. Obukhov for useful comments. They would %like to thank
%also FAPESP-Brazil, CNPq-Brazil and CAPES-Brazil for financial support.

\end{document}